\documentclass[sn-mathphys-num]{sn-jnl}

\usepackage{multirow}%
\usepackage{amsmath,amssymb,amsfonts}%
\usepackage{amsthm}%
\usepackage{mathrsfs}%
\usepackage[title]{appendix}%
\usepackage{xcolor}%
\usepackage{textcomp}%
\usepackage{manyfoot}%
\usepackage{booktabs}%
\usepackage{algorithm}%
\usepackage{algorithmicx}%
\usepackage{algpseudocode}%
\usepackage{listings}%

\usepackage{soul}
\usepackage{color}

\theoremstyle{thmstyleone}%
\theoremstyle{thmstyletwo}%

\theoremstyle{thmstylethree}%

\begin{document}

\title[On the Effectiveness of the `Follow-the-Sun' Strategy in Mitigating the Carbon Footprint of AI in Cloud Instances]{On the Effectiveness of the `Follow-the-Sun' Strategy in Mitigating the Carbon Footprint of AI in Cloud Instances}


\author*[1]{\fnm{Roberto} \sur{Vergallo}}\email{roberto.vergallo@unipegaso.it}

\author[2]{\fnm{Luís} \sur{Cruz}}\email{l.cruz@tudelft.nl}

\author[3]{\fnm{Alessio} \sur{Errico}}\email{alessio.errico@studenti.unisalento.it}

\author[3]{\fnm{Luca} \sur{Mainetti}}\email{luca.mainetti@unisalento.it}

\affil[1]{\orgdiv{Department of Information Sciences and Technologies}, \orgname{Pegaso Digital University}, \orgaddress{\street{Piazza Trieste e Trento 48}, \city{napoli}, \postcode{80132}, \state{Campania}, \country{Italy}}}

\affil[2]{\orgdiv{Department of Software Technology}, \orgname{Delft University of Technology}, \orgaddress{\street{Van Mourik Broekmanweg 6}, \city{Delft}, \postcode{2628 XE}, \state{Zuid-Holland}, \country{The Netherlands}}}

\affil[3]{\orgdiv{Department of Innovation Engineering}, \orgname{University of Salento}, \orgaddress{\street{via per Monteroni, 165}, \city{Lecce}, \postcode{73100}, \state{Apulia}, \country{Italy}}}


\abstract{``Follow-the-Sun'' (FtS) is a theoretical computational model aimed at minimizing the carbon footprint of computer workloads. It involves dynamically moving workloads to regions with cleaner energy sources as demand increases and energy production relies more on fossil fuels. With the significant power consumption of Artificial Intelligence (AI) being a subject of extensive debate, FtS is proposed as a strategy to mitigate the carbon footprint of training AI models. However, the literature lacks scientific evidence on the advantages of FtS to mitigate the carbon footprint of AI workloads. In this paper, we present the results of an experiment conducted in a partial synthetic scenario to address this research gap.

We benchmarked four AI algorithms in the anomaly detection domain and measured the differences in carbon emissions in four cases: no strategy, FtS, and two strategies previously introduced in the state of the art, namely Flexible Start and Pause and Resume. To conduct our experiment, we utilized historical carbon intensity data from the year 2021 for seven European cities. Our results demonstrate that the FtS strategy not only achieves average reductions of up to 14.6\% in carbon emissions (with peaks of 16.3\%) but also helps in preserving the time needed for training.}




\maketitle

\section{Introduction}
\label{sec:introduction}
The environmental impact of technological development, especially in the world of software, has become a topic of increasing relevance and concern in recent years~\cite{trefethen2013energy}. Technological advancements and the widespread use of software applications and devices have resulted in a range of often underestimated environmental consequences. While modern technologies have undoubtedly improved our lives in many ways, they are also responsible of generating a significant amount of Greenhouse Gas (GHG) emissions, energy consumption, and waste production.

Specifically, the growing dependence on Artificial Intelligence (AI) is not without consequences. The substantial computational resources required for training Deep Neural Networks (DNN) and running complex algorithms have led to a significant increase in CO\textsubscript{2} emissions. Recent research has shown the environmental impact of AI, as in the case of GPT-2, where emissions are five times greater than those of a car throughout its entire lifecycle~\cite{strubell2020energy}. The AI community is recognizing the urgency of addressing the environmental impact, and efforts are underway to develop ``green AI'' solutions. The goal is to develop algorithms and infrastructure that are energy-efficient, significantly reducing carbon emissions associated with AI training and execution. Particularly, the use of Cloud technologies represents an opportunity for managing the environmental impact. Having data centers in different geographical regions means that each instance, by consuming energy from its grid-region, has different carbon emissions. This approach aligns with a historical period where the `Follow-the-Sun' (FtS)~\cite{fts} model is proposed for various problems, both for cross-cloud application migration \cite{shen2016follow} but also for example as a case of Global Software Development~\cite{kroll2018empirical, santos2015using, serwinski2019identification}, all with the goal of maximizing efficiency through the involvement of different geographic areas.

However, the implementation of FtS is far from trivial because it requires a significantly more complex IT architecture. This architecture involves multiple GPU servers spread all over the globe, which need to be effectively orchestrated. Such complexity translates in higher implementation costs. Moreover, such IT assets, even if paused in their usage while training is performed in other zones of the world, have costs for their reservation in the case of Cloud providers, that should be taken into account. Finally, moving data across national borders follows strict regulations.
Ensuring legal compliance and implementing adequate data protection measures to meet these regulations incurs additional costs.

Before organizations begin the path of reorganizing their IT infrastructure to implement FtS, scientific evidence on the effectiveness of FtS is needed to help managers make informed decisions. However, the state of the art is still lacking scientific evidence on the effectiveness of the FtS strategy applied to AI training workloads. The main purpose of this work is to measure how FtS can contribute to improving the carbon footprint of AI training, and ultimately determine whether the respective costs and efforts are justified.

While AI contributes to carbon emissions in both training and inference phases, we deliberately chose to focus only o                of massive carbon emissions, the related environmental cost falls in the production phase of software life cycle. Training instead falls in the development phase. Carbon emissions of development are new\footnote{The rebound effect is also applicable for the production phase of AI, as for many other technologies bringing efficiency improvements}.

For the sake of completeness, this work also compares the performance of FtS with two state-of-the-art strategies, that use time flexibility as strategy to minimise the carbon footprint of AI workloads -- namely, `Flexible start' and `Pause and Resume'~\cite{dodge2022measuring}.

To summarize, contributions of this paper are the following:

\begin{itemize}
    \item We perform a retrospective analysis to understand how carbon emissions of AI training workloads change if performed with or without FtS;
    \item We compare FtS with two existing carbon-aware training strategies to show the extent to which FtS performs better in terms of improving the carbon footprint of AI training while preserving training time;
    \item We provide source code, dataset and results to encourage open research.
\end{itemize}

The source code is organized in two Python projects, each of them is available on GitHub. The first project\footnote{\url{https://github.com/softengunisalento/AI_Train_Workload}} is aimed at running the AI training workloads and capture energy consumption. The aim of the second project\footnote{\url{https://github.com/softengunisalento/AI_TrainingStartegyLauncher}} is to launch the carbon-aware training strategies and to collect results. The historical marginal carbon emission data and the AI dataset are available upon request. Moreover, the source code of one of the selected AI algorithms (HF-SCA) is not disclosed for industrial purposes.

The remainder of the paper is organized as follows: in Section \ref{sec:related-works}, we provide an overview on recent advancements in research concerning carbon-aware training strategies, specifically Flexible Start and Pause and Resume. Section \ref{sec:proposed-models} introduces the research methodology, including the proposed FtS model and its variants. The main findings of the experiment and the evaluation of the FtS strategy are presented in Section \ref{sec:experimental-results}. Section \ref{sec:discussion} contains the discussion. Final remarks and open research challenges are reported in Section \ref{sec:conclusions}.

\section{Background}\label{sec:related-works}
The studies in the field of AI has experienced significant growth. This development has led to increasing attention being paid to side effects such as its environmental impact, resulting in further expansion of research in this area. A recent mapping study on green AI~\cite{verdecchia2023systematic} reported that publications in this field have increased by 76\% since 2020. Despite being a quite new research topic (with the first paper on Green AI being published in 2015~\cite{kung2015}), the Green AI research field seems to have positioned and consolidated itself quite quickly within AI research communities. However, the mapping study observes that only 18\% of the 98 primary studies focus on the carbon footprint~\cite{wiedmann2008definition}, with the majority (74\%) focusing on energy efficiency~\cite{verdecchia2021green}, and a few (8\%) on the ecological footprint indicator~\cite{matuvstik2021footprint} which has been often criticized \cite{van1999spatial}\cite{kitzes2009research}\cite{giampietro2014footprints}\cite{van2014ecological}\cite{van2014ecological}. Despite energy efficiency metrics are relatively easy to compute, they do not directly reflects on the impact on the global warming issue, because energy is a variable mix of green, clean and fossil sources. Moreover, energy efficiency and carbon efficiency are not alternative but sequential models: after we make an algorithm energy efficient, we can make it carbon efficient using some carbon-aware training strategies which for example take into account the carbon intensity (measured in gCO\textsubscript{2}eq/kWh) of the region where the server is located. So the mapping study suggests that research efforts should be encouraged to increase our knowledge in carbon-aware AI deployments.

The literature provides only one paper focused on deploying AI algorithms in Cloud instances and monitoring their carbon emissions, which is Dodge et.al.~\cite{dodge2022measuring}. In fact, the replication package provided in Verdecchia et.al.~\cite{verdecchia2023systematic} and a thorough recursive bidirectional snowballing process~\cite{wohlin2014guidelines} confirmed that at the time of writing (June 2024) their carbon-aware training strategies represent the state of art in this field. The authors propose two training strategies for Machine Learning (ML) algorithms based on the use of Cloud architectures. As expressed by the authors themselves, they introduce the first tool to estimate the real-time CO\textsubscript{2} emissions impact of instances on a Cloud computing platform. This post-hoc study investigates what the CO\textsubscript{2} emissions would have been if a given workload had been performed under different conditions and in different ways. The considered workloads are several model training concerning NLP and Computer vision tasks. For each training they calculated the operational emissions by making use of:
\begin{itemize}
    \item GPUs energy consumption tracked $5m$;
    \item historical data of local-marginal emissions supplied by WattTime\footnote{https://www.watttime.org/} with a $5m$ granularity.
\end{itemize}

The two strategies benchmarked by Dodge et. al. are named 'Flexible Start' and 'Pause \& resume'. Flexible Start (FS) is the first strategy proposed. Given a time window, this algorithm wants to find out the best starting time in term of carbon emissions. Considering all the possible starting times in the time window, the strategy consists in picking up the one that would produce the lowest emissions. This strategy wants the workload to run until completion. Pause and Resume (PaR) instead implies that the job could be stopped at certain point of the run and then resume. Assuming that stopping a job doesn't requires additional cost, the algorithm, in a given time window, selects all the lowest marginal emission $5m$ intervals until the length of the workload is covered. Once the intervals are found, then the corresponding emissions are computed.

The strategies were compared by taking the marginal carbon intensity data throughout 2020. While the implementation of the strategies is quite simple, the results are very interesting, as they are remarkably efficient in complementary cases.
\begin{itemize}
    \item \textbf{Flexible Start}\newline
    It has been shown to be particularly efficient for short duration workloads, as in the case of Densnet: the best emissions reduction reaches 80\% in the West US region. In contrast, for jobs longer than one day the reductions are less significant. In fact, the same strategy applied to a 6 billion parameters transformer lead to a reduction lower than 2\%. This may be explained by the fact that a shorter job is less subject to the variability of marginal emissions during the time window.
    \item \textbf{Pause and Resume}\newline
   It turns out to have considerable reductions on those regions that have wide variability of marginal emissions during a single day and when a workload is longer than a day. This strategy applied to Densnet brought a small reduction, lesser than 10\%. Contrariwise, applied to a 6 billion parameters transformer the reduction reaches almost 30\%.
\end{itemize}

It should be kept in mind that using historical data leads to get the best results, because in real scenarios the strategies would use forecasted data. The related emission reductions turn out to be a lower bounds for realistic scenarios. 

\subsection{Motivations}
Dodge et.al. have proposed interesting and valid solutions, involving the technology that paradoxically contribute to massive carbon emissions: the Cloud. Their work has highlighted the potential reduction in CO$_{2}$ emissions by performing training during the best time of day or distributing it across time intervals with lower carbon intensity. Their results, aggregated across different regions and periods of the year, emphasize the importance of time and location as determining factors for pollution. However, their solution, even though the results were averaged across multiple regions, solely relies on time as a resource.
This evidences two important gaps:
\begin{itemize}
    \item state-of-the-art solutions leverages time without considering the fact that time can also be a valuable resource, as it translates into costs for the industry;
    \item emissions depend not only on the time of day but also on the grid-region, but this possibility has not been explored.
\end{itemize}
The aim of the next sections is to face the previously listed research gaps by benchmarking the FtS carbon-aware training strategy.

\section{Study design}\label{sec:proposed-models}
This section report on the research methodology we adopted by formulating the research questions, detailing the experiment, presenting the benchmarks we selected, describing the employed metrics, and discussing the implementations.

\subsection{Research questions}\label{sec:rq}
The research questions we address in this work are the following:
\begin{itemize}
    \item \textbf{RQ1} \textit{How is FtS effective in reducing the carbon footprint of training AI algorithms?} Moving AI training workload across Cloud instances by checking for better carbon intensity is supposed to save carbon emissions. With this research question we want to quantify the benefit of FtS executing a benchmark consisting in different kinds of AI workloads.
    \item \textbf{RQ2} \textit{How does FtS perform with respect to the state-of-the-art training strategies?} FS and PaR are the two existing strategies. To address this research question we re-implemented such strategies and measure them over the same benchmark.
    \item \textbf{RQ3} \textit{How is FtS effective in saving training time with respect to state-of-the-art?} We anticipate that FtS can lead to time savings because it does not need to wait for better carbon intensity.
\end{itemize}

\subsection{Design of the experiment} \label{sec:design}
\begin{figure}
\centering
\includegraphics[width=0.75\linewidth]{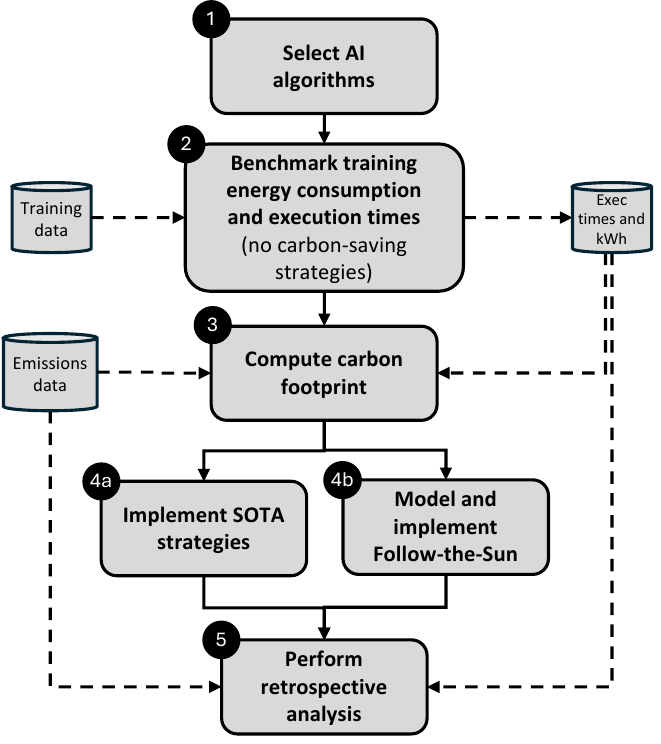}
  \caption{ Research methodology adopted in this paper.}
  \label{fig:research_method}
  \end{figure}

In this work, we have conducted a retrospective analysis to understand what the carbon emissions could be if a training workload was executed in one way rather than another. Figure~\ref{fig:research_method} reports on the research methodology we adopt in this paper. Each step of the experiment is described afterwards.

Step 1 consisted in selecting the AI algorithms used as a benchmark for the carbon saving strategies. Such benchmark was needed to compare the different training strategies in an partially synthetic scenario. The analysis was conducted with the same objective as Dodge et.al.~\cite{dodge2022measuring}: implementing the training strategy by first understanding the consumption patterns in the best-case conditions, and then providing users with the opportunity to use this strategy by using forecast data.

To know how much carbon each strategy emits to the atmosphere, first we measured the energy consumed by the AI algorithm on the same machine, sampled every $5m$ (step 2). The power consumption was recorded using the Python library $CodeCarbon$\footnote{https://codecarbon.io/}. CodeCarbon library provides out of the box, in addition to energy measurements, also estimates regarding $CO_2$ emissions. We instead used the marginal carbon emissions dataset provided by the research partner WattTime to map the energy consumption to historical carbon emission (step 3). WattTime provided data for the year 2021 for 7 European cities (Milan, Paris, Frankfurt, Zaragoza, London, Dublin, and Stockholm) within the context of the Green Software Foundation\footnote{https://greensoftware.foundation/} mutual agreement. Such cities were selected because they host AWS regions, so they represent a realistic scenario for our research. We selected only European cities in order not to violate the Global Data Protection Regulation (GDPR).

Once the energy consumption data and the marginal emissions were obtained, we reimplemented the code of FS and PaR from state-of-the-art (step 4a), and we implemented the FtS strategy from scratch (step 4b). Finally, we performed the retrospective analysis to compare the different strategies (step 5). Specifically, all strategies were executed under the same conditions for the considered year, 2021, as well as regions and days for each month. The idea is to calculate the reductions in carbon emissions for each strategy by averaging the results over the entire year of 2021. For each month, we selected 6 equal starting times corresponding to 6 days. This choice was made to accurately describe the average emissions for each month. This approach aligns with the methodology used by Dodge et.al.~\cite{dodge2022measuring} to summarize annual emissions. To provide more precise results, in the case of state-of-the-art strategies, we averaged not only for the entire year of 2021 but also across all regions.

The benchmark also considers the average number of workload transfers that the FtS would have over the year. 

Furthermore, we compared the average duration of the workloads for the different strategies. We used this metric to understand the extent to which state-of-the-art strategies can save training time. All strategies were subjected to the same time windows, calculated using hour-based values (more on this in Section~\ref{sec:res_eval}).

We carefully report all these metrics to show the average performance of the strategies for each individual training, particularly to summarize the average carbon reductions and average time extensions.

\subsection{Follow-the-Sun}
The name seems to refer specifically to solar energy, but actually it describes a strategy that aims to harness all the renewable energy sources. It is a name that effectively captures the basic functioning of the strategy: moving in pursuit of the cleanest energy sources, much like a sunflower follows the movement of the sun for its energy needs. 

It is based on the Cloud technology for two reasons: 
\begin{itemize}
    \item as mentioned earlier, Cloud instances have enabled access to the necessary hardware resources for AI training workloads;
    \item Cloud instances can be chosen by the user based on their region.
\end{itemize}  

This implies the ability to use multiple instances as part of a network where the workload is transferred to the optimal region whenever emissions exceed a certain threshold.

To enrich our analysis of FtS, and to leverage the advantages brought by state-of-the-art strategies, we analize the $Static-Start \ FtS$ (ssFtS) and the $Flexible-Start \ FtS$ (fsFtS) variants, which differs in their starting times because, as the name suggests, the second version leverages the underlying principle of FS from state-of-the-art.

Finally, for each of these versions, we evaluate two workload transfer modes:
\begin{itemize}
    \item  Upstream transfer, i.e. move the dataset to each Cloud instance before the training begins. Main disadvantage: distributing to all instances can result in transferring the workload to regions that may never be used.
    \item In-training transfer, i.e. transfer the dataset during training where the energy is greener. Main disadvantage: transfer times may be not negligible.
\end{itemize}

In total, we measured four versions of FtS:
\begin{enumerate}
    \item Upstream Static-Start Follow-the-Sun
    \item In-traning Static-Start Follow-the-Sun
    \item Upstream Flexible-Start Follow-the-Sun
    \item In-traning Flexible-Start Follow-the-Sun
\end{enumerate}

For the readers' convenience, Table \ref{tab:acronyms} lists the acronyms for all the strategies benchmarked in this paper.

\begin{table}[]
\begin{tabular}{| p{0.1\linewidth} | p{0.2\linewidth} | p{0.5\linewidth} |}
\hline
Acronym & Extended name                 & Purpose                                                                                     \\ \hline
FS      & Flexible Start                & Wait for better Carbon Intensity before launching the training                              \\ \hline
PaR     & Pause \& Resume               & Pause training when Carbon Intensity is too high                                            \\ \hline
FtS     & Follow-the-Sun                & Move workload geographically as the Carbon Intensity get too high. Does not wait nor pause. \\ \hline
ssFtS   & Static-Start Follow-the-Sun   & Same as FtS                                                                                 \\ \hline
fsFtS   & Flexible-Start Follow-the-Sun & Same as FtS but wait if Carbon Intensity is too high (as FS)                                \\ \hline
\end{tabular}
\caption{Recap of the acronym used in this paper to refer to the benchmarked carbon-aware training strategies.}
\label{tab:acronyms}
\end{table}

\subsubsection{Static-Start Follow-the-Sun}

This strategy takes three input parameters: 
\begin{itemize}
    \item Starting time. it represents the point in time when the training should start.
    \item Reference Region. It is the initial region where the Cloud instance is located and where dataset and workloads are loaded. It is the first to receive the workload in the case of in-training transfer.
    \item Checking time. Describes how often the strategy checks whether to continue training on the current region or move to another one.
\end{itemize}
The workload starts at the specified starting time, and at each checking time it possibly moves to the greenest region to continue with the training. In more technical terms, starting from the given input starting time, the workload is split into $k$ slots based on the selected checking time. Each slot corresponds to a training segment that is executed on the region with the least environmental impact. Since we are focusing on operational emissions (scalar O) only, partial emissions are calculated for each slot by taking the dot product between the corresponding subset of energy consumption (row vector $\mathbf{E}$ with dimensions $(1, k)$) and the subset of marginal emissions for each region. The regions with minimal dot product values is designated to execute the corresponding training segment during that time slot.
In mathematical terms, we can write: 
\begin{equation}
    O_{ssFtS}=\mathbf{E}\cdot \mathbf{I}_{R_{best}}
\end{equation}
$\mathbf{I}_{R_{best}}$ represents the column vector (with dimensions $(k, 1)$) containing the marginal emissions values of the greenest regions for the given starting time. In particular, it is composed of the concatenation of $k$ sets of values.

\begin{equation}
\mathbf{I}_{R_{best}}=
    \begin{bmatrix}
        I_{1,r_{best}} \\
        I_{2,r_{best}} \\
        \vdots \\
        I_{\gamma,r_{best}} \\
        \vdots \\
        I_{k,r_{best}}
    \end{bmatrix}
\end{equation}

$I_{\gamma,r_{best}}$ refers to the marginal emissions values associated with the $\gamma-th$ time slot of the region that would result in the minimum partial operational emissions. Therefore, by calculating the partial dot product between this set and the corresponding energy consumption set, the minimum partial operational emission is obtained, which depends on the region $r$. Thus, the best marginal emissions at the time specified by the $\gamma$ index are:
\begin{equation}
I_{\gamma,r_{best}}= \arg \min_{r}  (I_{\gamma,r} \cdot E_{\gamma} )
\end{equation}
where $1\leq \gamma \leq k$, and $r$ represents the index for the regions.

Therefore, $E_{\gamma}$ is the $\gamma-th$ value in $\mathbf{E}$ corresponding to the time slot $\gamma$:
\begin{equation}
    \mathbf{E}=[E_{1}|E_{2}...|E_{\gamma}...|E_{k}]
\end{equation}

\subsubsection{Flexible-Start Follow-the-Sun}
The input parameters are the same as those for the static version, with the addition of one more: the ending time.
This parameter defines a time window within which the training can be optimally scheduled. 

As the name suggests, this strategy is a hybrid of FS and ssFtS. Specifically, it is subjected to the main characteristic of both strategies:
\begin{itemize}
    \item starting at the time that minimizes emissions without interrupting the execution;
    \item conducting the training in the most environmentally friendly regions.
\end{itemize} 

The functioning is similar to that of the static version, except that this time the division into slots applies to all the possible starting times. Since this strategy waits for the best moment to start, FS represents a lower bound, and it has a gain equal to the static version in the worst-case scenario.
\begin{equation}
    O_{fsFtS}= \min_{i \in [s, e]} (\mathbf{E}\cdot \mathbf{I}_{R_{best}}^{i}) \quad i+k\leq e
\end{equation}

$\mathbf{I}_{R_{best}}^{i}$ is the result of concatenating the portions of marginal emissions corresponding to the $\gamma-th$ time slot associated with the $i-th$ starting time.

\begin{equation}
\mathbf{I}_{R_{best}}^{i}=
    \begin{bmatrix}
        I_{1,r_{best}}^{i} \\
        I_{2,r_{best}}^{i} \\
        \vdots \\
        I_{\gamma,r_{best}}^{i} \\
        \vdots \\
        I_{k,r_{best}}^{i}
    \end{bmatrix}
\end{equation}

 The generic $I_{\gamma,r_{best}}^{i}$ is the best portion of marginal emissions corresponding to the $\gamma-th$ time slot relative to the best region. Specifically, it represents the subset of marginal emission values that, when multiplied by the corresponding subset of power consumptions, yields the lowest operational emissions value.
\begin{equation} 
I_{\gamma,r_{best}}^{i}= \arg \min_{r}  (I_{\gamma,r}^{i} \cdot E_{\gamma} )
\end{equation}

\subsection{Test environment}
The machine used to benchmark the AI algorithms is CINECA's DGX, a NVIDIA A100 accelerated system available to the Italian public researchers from January 2021. The machine is equipped with 1 NVIDIA A100-SXM4-40GB GPU, 1 AMD EPYC 7742 64-Core Processor and 1TB RAM. CINECA consortium is the largest Italian computing centre. Python version is 3.8.10.

\subsection{AI workloads and dataset} \label{AIWL}
AI algorithms have been selected from the field of Anomaly Detection (AD), particularly $fraud \ detection$, because of the availability of a large labelled dataset of real banking transactions provided by a mid-sized Italian bank. The choice of AD was made also to demonstrate that classical and simpler ML workloads have a not negligible environmental impact, so not only heavier workloads like training Large Language Models (LLP), NLP and computer vision models are harmful for the environment. The dataset is structured as follows. It is a tabular dataset with 32 features (columns), plus the column for the ground truth which is a boolean. The features are anonymized and scaled through both standard and min-max scalers. The number of rows is 4.5 millions. The size of the dataset is $2GB$.

The domain of AD ranges from unsupervised to supervised learning, and even includes deep learning algorithms. A brief review of the considered workload is provided below:
\begin{itemize}
    \item Isolation Forest (IF). It is an algorithm used in AD to identify anomalies within a dataset~\cite{liu2008isolation}. It is based on an unsupervised learning approach and focuses on the idea that anomalies are rarer and therefore more ``isolated'' than normal instances.
    \item Support Vector Machine (SVM). It is a ML algorithm used for classification and regression problems~\cite{hearst1998support}. It is a supervised learning model that can be used to separate data examples into different classes.
    \item Autoencoder (AE). It is a neural network architecture commonly used in AD~\cite{bank2020autoencoders}. It falls under the category of unsupervised learning, as it learns to reconstruct the input data without relying on labelled examples.
    \item HF-SCA. It is an unsupervised DL algorithm~\cite{distante2022hf} based on a U-Net, a specific neural network whose architecture closely resembles the letter ``U''. This structure is designed in such a way that information, starting from the top left of the ``U'', is progressively compressed through a process of downsampling using convolutional layers and pooling. Once it reaches the bottom point, the information then travels back up towards the top right of the ``U'', reconstructing the data through an upsampling process.
\end{itemize}

Once the training was set up on the DGX at CINECA, the training processes were executed. Specifically, for each algorithm, a grid search was launched, which extended for several hours depending on the AD algorithm. The grid search was the considered hyperparameters fine tuning method both to simulate the worst-case scenario (in terms of energy consumption), and also to have significant workloads. The grid search was based on the Area Under the ROC Curve (AUC) score metric, as the AUC score is preferred in AD.

\subsection{Workload transfer cost} \label{sec:cost}
We defined a partially synthetic scenario to fairly measure the performances of all the strategies -- including the four versions of FtS -- in a controlled environment. While energy consumption of the selected AI algorithms was measured on a real machine, and the dataset of historical carbon emission data was real, there are some assumptions that we made to delimit our partially synthetic scenario.

In the previous subsection, we declared the size of the dataset used for training. However, the guidelines provided by the Green Software Foundation suggest compressing textual data, as the transfer would have a lesser impact. Therefore, the dataset size is reduced from $2GB$ to approximately $0.320GB$. Given the high efficiency of the interconnections among data centers, we consider the transfer times as negligible. This could be not necessarily true for large dataset and would imply having time slots dependent on the transfer time. Furthermore, files containing training state information typically weigh less than $100MB$, which, once compressed, have sizes that are insignificant compared to the calculation of energy consumption, so we neglect them as well.

In this work we consider an energy consumption of $0.023kWh$ per GigaByte of data transferred, as suggested by Malmodin and Lunden \cite{malmodin2016energy}.

Moreover, stopping and resuming a job comes at a cost -- in terms of time and energy -- that we neglected. Additionally, since training of the neural networks is a stochastic procedure, it depends on the order of operations performed partially in parallel. We do not know if the results obtained by training the same algorithm with different strategies are identical. Stopping the training process and moving the environment to another computing node could change the convergence of the loss function during the training, and the training could hence be longer. Nevertheless, in our partial synthetic scenario we ideally assume that all the training workloads we considered can be stopped and resumed in another place.

Finally, in our partial synthetic scenario we assume that all the involved computing locations are free and available to receive new tasks. This is necessarily not true in the real world.

\section{Results}\label{sec:experimental-results}
\begin{figure}
\centering
\includegraphics[width=0.75\linewidth]{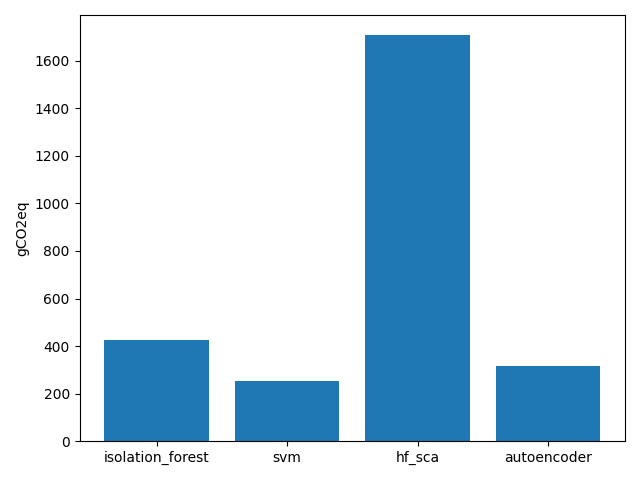}
  \caption{ Average emissions of the trainings during the year 2021 across different regions.}
  \label{fig:no_eco_em}
  \end{figure}

Before presenting the results for each strategy, it is important to provide an overview of the emissions produced by the four algorithms with no strategy, to serve as a baseline. In this case, energy data was mapped with historical data on local marginal emissions, calculating the average total emissions for the entire year 2021, applied to the 7 regions mentioned in Section \ref{sec:design} (Figure \ref{fig:no_eco_em}).

The IF execution lasted approximately $4.25h$, resulting in a consumption of $0.825kWh$ of electrical power. The SVM run approximately for $2.5h$, resulting in energy consumption of $0.493kWh$. The AE workload lasted approximately $3.5h$, with consequent energy consumption amounting to $0.615kWh$. The HF-SCA workload is the longest among all, with a duration of $16h$, mainly because the grid search involved the training of 108 models, corresponding to the same number of hyperparameters combinations (three values for learning rates $\times$ 6 values for memory items $\times$ 6 values for number of hidden layers). Being so extensive in time, it also resulted in the highest energy consumption: $3.310kWh$.

The heaviest workload was the HF-SCA algorithm. Its emissions exceed $1,600gCO_2eq$, comparable to emissions of CO$_{2}$ per litre of fuel consumed by a car. For the sake of completeness, in Table \ref{tab:auc} the AUC scores achieved by the algorithms are reported. We report this metric for a specific reason: our aim is to provide an overview in terms of consumption and accuracy to guide a decision maker in making an informed choice regarding a training process, i.e. to find a balance between model accuracy and its environmental impact. Among the different workloads, HF-SCA stands out significantly, but as reported before, the idea is to consider multiple workloads and analyze their performance and consumption to provide information needed to take decisions. This type of proactive decision-making already mitigates emissions, which is further reduced by the strategy benchmarked in this work.

\begin{table}[t]
    \centering
    \small
    \begin{tabular}{|c|c|c|c|}
        \hline
        Isolation Forest & SVM & Autoencoder & HF-SCA\\
        \hline
        0.56 & 0.51 & 0.73 & 0.97\\
        \hline
    \end{tabular}
    \caption{AUC score for each model}
    \label{tab:auc}
\end{table}

\subsection{Results and evaluation} \label{sec:res_eval}
In this section, we show the average reductions in carbon emissions throughout the year 2021 for all the implemented strategies. As already stated in Section \ref{sec:proposed-models}, the analysis considered 6 days per month. We adopted this approach to closely align with the \textit{modus operandi} of Dodge~et~al.~\cite{dodge2022measuring}, hence ensuring a reliable and fair comparison.

Furthermore, this evaluation involves calculating the reductions based on the variation of key input parameters of the strategies, particularly:
\begin{itemize}
    \item The time window is calculated by adding each value in the set [6, 12, 18, 24] -- which we refer to as the hours-set -- to the length of the workload.
    \item The checking time falls in the set [15, 30, 60, 120] (in minutes).
\end{itemize}

Tables \ref{tab:hoursset1}-\ref{tab:hoursset4} show the variation in carbon emissions for the four FtS variants and for each combination of checking time and time window parameters.

\begin{table}[ht]
  \begin{minipage}{0.45\linewidth}
    \centering
    \small
    \setlength\tabcolsep{9pt} 
\begin{tabular}{|c|cccc|}
\hline
\multirow{2}{*}{Workload} & \multicolumn{4}{c|}{Checking time}                                                   \\
                          & \multicolumn{1}{c|}{15}  & \multicolumn{1}{c|}{30}  & \multicolumn{1}{c|}{60}  & 120 \\ \hline
IF                        & \multicolumn{1}{c|}{6.7} & \multicolumn{1}{c|}{6.7} & \multicolumn{1}{c|}{6.7} & 6.3  \\ \hline
SVM                       & \multicolumn{1}{c|}{6.0} & \multicolumn{1}{c|}{6.0} & \multicolumn{1}{c|}{6.0} & 5.9 \\ \hline
AE                        & \multicolumn{1}{c|}{6.3} & \multicolumn{1}{c|}{6.3} & \multicolumn{1}{c|}{6.2} & 6.2 \\  \hline
HF-SCA                    & \multicolumn{1}{c|}{5.5} & \multicolumn{1}{c|}{5.3} & \multicolumn{1}{c|}{5.3} & 5.2 \\  \hline
\end{tabular}

\caption{Emission reduction (in percentage) for hours-set for upstream ssFtS} \label{tab:hoursset1}
  \end{minipage}\hfill
  \begin{minipage}{0.45\linewidth}
\small
\setlength\tabcolsep{9pt} 
\begin{tabular}{|c|cccc|}
\hline
Workload & \multicolumn{4}{c|}{Checking time} \\ 
         & \multicolumn{1}{c|}{15} & \multicolumn{1}{c|}{30} & \multicolumn{1}{c|}{60} & 120 \\ \hline
IF       & \multicolumn{1}{c|}{6.0} & \multicolumn{1}{c|}{6.0} & \multicolumn{1}{c|}{6.0} & 6.0 \\ \hline
SVM      & \multicolumn{1}{c|}{5.8} & \multicolumn{1}{c|}{5.8} & \multicolumn{1}{c|}{5.9} & 5.8 \\ \hline
AE       & \multicolumn{1}{c|}{5.5} & \multicolumn{1}{c|}{6.0} & \multicolumn{1}{c|}{5.8} & 6.1 \\ \hline
HF-SCA   & \multicolumn{1}{c|}{5.0} & \multicolumn{1}{c|}{5.0} & \multicolumn{1}{c|}{5.0} & 4.8 \\ \hline
\end{tabular}

\caption{Emission reduction (in percentage) for hours-set for in-training ssFtS} \label{tab:hoursset2}
  \end{minipage}
\end{table}

\begin{table}[ht]
  \begin{minipage}{0.45\linewidth}
    \centering
    \small
\setlength\tabcolsep{3.1pt} 
\begin{tabular}{|c|c|cccc|}
\hline
\multirow{2}{*}{Workload} & \multirow{2}{*}{\begin{tabular}[c]{@{}c@{}}Time\\ window\end{tabular}} & \multicolumn{4}{c|}{Checking time}                                                   \\
                          &                                                                        & \multicolumn{1}{c|}{15}  & \multicolumn{1}{c|}{30}  & \multicolumn{1}{c|}{60}  & 120 \\ \hline
\multirow{4}{*}{IF}       & 6                                                                      & \multicolumn{1}{c|}{7.8} & \multicolumn{1}{c|}{7.8} & \multicolumn{1}{c|}{7.8} & 7.8 \\
                          & 12                                                                     & \multicolumn{1}{c|}{9.0} & \multicolumn{1}{c|}{9.0} & \multicolumn{1}{c|}{8.8} & 8.5 \\ 
                          & 18                                                                     & \multicolumn{1}{c|}{12.0} & \multicolumn{1}{c|}{12.0} & \multicolumn{1}{c|}{11.8} & 11.5 \\
                          & 24                                                                     & \multicolumn{1}{c|}{15.7} & \multicolumn{1}{c|}{15.7} & \multicolumn{1}{c|}{15.5} & 15 \\ \hline
\multirow{4}{*}{SVM}      & 6                                                                      & \multicolumn{1}{c|}{7.8} & \multicolumn{1}{c|}{8.0} & \multicolumn{1}{c|}{7.9} & 7.7 \\
                          & 12                                                                     & \multicolumn{1}{c|}{9.8} & \multicolumn{1}{c|}{9.8} & \multicolumn{1}{c|}{9.8} & 9.3 \\ 
                          & 18                                                                     & \multicolumn{1}{c|}{11.7} & \multicolumn{1}{c|}{11.5} & \multicolumn{1}{c|}{11.0} & 11.0 \\ 
                          & 24                                                                     & \multicolumn{1}{c|}{\textbf{16.3}} & \multicolumn{1}{c|}{\textbf{16.3}} & \multicolumn{1}{c|}{16.2} & 15.8 \\ \hline
\multirow{4}{*}{AE}       & 6                                                                      & \multicolumn{1}{c|}{8.0} & \multicolumn{1}{c|}{7.8} & \multicolumn{1}{c|}{7.8} & 7.7 \\
                          & 12                                                                     & \multicolumn{1}{c|}{9.7} & \multicolumn{1}{c|}{9.3} & \multicolumn{1}{c|}{9.0} & 9.0 \\ 
                          & 18                                                                   & \multicolumn{1}{c|}{11.8} & \multicolumn{1}{c|}{11.3} & \multicolumn{1}{c|}{11.0} & 10.3 \\ 
                          & 24                                                                     & \multicolumn{1}{c|}{\textbf{16.3}} & \multicolumn{1}{c|}{\textbf{16.3}} & \multicolumn{1}{c|}{16.2} & 15.8 \\ \hline
\multirow{4}{*}{HF-SCA}   & 6                                                                      & \multicolumn{1}{c|}{9.0} & \multicolumn{1}{c|}{9.0} & \multicolumn{1}{c|}{8.8} & 8.6 \\
                          & 12                                                                     & \multicolumn{1}{c|}{9.2} & \multicolumn{1}{c|}{9.4} & \multicolumn{1}{c|}{9.0} & 8.7 \\ 
                          & 18                                                                     & \multicolumn{1}{c|}{9.7} & \multicolumn{1}{c|}{9.7} & \multicolumn{1}{c|}{9.6} & 9.3 \\ 
                          & 24                                                                     & \multicolumn{1}{c|}{10.0} & \multicolumn{1}{c|}{10.0} & \multicolumn{1}{c|}{9.9} & 9.7 \\ \hline
\end{tabular}
\caption{Emission reduction (in percentage) for hours-set for upstream fsFtS} \label{tab:hoursset3}
  \end{minipage}\hfill
  \begin{minipage}{0.45\linewidth}
\small
\setlength\tabcolsep{3.1pt} 
\begin{tabular}{|c|c|cccc|}
\hline
\multirow{2}{*}{Workload} & \multirow{2}{*}{\begin{tabular}[c]{@{}c@{}}Time\\ window\end{tabular}} & \multicolumn{4}{c|}{Checking time}                                                   \\
                          &                                                                        & \multicolumn{1}{c|}{15}  & \multicolumn{1}{c|}{30}  & \multicolumn{1}{c|}{60}  & 120 \\ \hline
\multirow{4}{*}{IF}       & 6                                                                      & \multicolumn{1}{c|}{7.0} & \multicolumn{1}{c|}{7.0} & \multicolumn{1}{c|}{7.0} & 7.0 \\ 
                          & 12                                                                     & \multicolumn{1}{c|}{8.5} & \multicolumn{1}{c|}{8.5} & \multicolumn{1}{c|}{8.0} & 8.0 \\ 
                          & 18                                                                     & \multicolumn{1}{c|}{11.0} & \multicolumn{1}{c|}{11.0} & \multicolumn{1}{c|}{11.0} & 11.0 \\ 
                          & 24                                                                     & \multicolumn{1}{c|}{15.0} & \multicolumn{1}{c|}{15.0} & \multicolumn{1}{c|}{15.0} & 14.5 \\ \hline
\multirow{4}{*}{SVM}      & 6                                                                      & \multicolumn{1}{c|}{7.3} & \multicolumn{1}{c|}{7.3} & \multicolumn{1}{c|}{7.7} & 7.6 \\
                          & 12                                                                     & \multicolumn{1}{c|}{9.0} & \multicolumn{1}{c|}{9.0} & \multicolumn{1}{c|}{9.0} & 9.0 \\ 
                          & 18                                                                     & \multicolumn{1}{c|}{11.0} & \multicolumn{1}{c|}{11.6} & \multicolumn{1}{c|}{11.0} & 10.8 \\ 
                          & 24                                                                     & \multicolumn{1}{c|}{16.0} & \multicolumn{1}{c|}{16.0} & \multicolumn{1}{c|}{16.0} & 15.5 \\ \hline
\multirow{4}{*}{AE}       & 6                                                                      & \multicolumn{1}{c|}{\textbf{6.2}} & \multicolumn{1}{c|}{7.3} & \multicolumn{1}{c|}{7.6} & 7.6 \\
                          & 12                                                                     & \multicolumn{1}{c|}{\textbf{8.0}} & \multicolumn{1}{c|}{8.8} & \multicolumn{1}{c|}{7.0} & 8.8 \\ 
                          & 18                                                                     & \multicolumn{1}{c|}{\textbf{10.0}} & \multicolumn{1}{c|}{10.2} & \multicolumn{1}{c|}{10.2} & 10.1 \\
                          & 24                                                                     & \multicolumn{1}{c|}{\textbf{15.5}} & \multicolumn{1}{c|}{15.8} & \multicolumn{1}{c|}{16} & 15.6 \\ \hline
\multirow{4}{*}{HF-SCA}   & 6                                                                      & \multicolumn{1}{c|}{8.5} & \multicolumn{1}{c|}{8.7} & \multicolumn{1}{c|}{5.3} & 5.0 \\
                          & 12                                                                     & \multicolumn{1}{c|}{8.8} & \multicolumn{1}{c|}{9.0} & \multicolumn{1}{c|}{5.3} & 5.0 \\ 
                          & 18                                                                     & \multicolumn{1}{c|}{9.2} & \multicolumn{1}{c|}{9.5} & \multicolumn{1}{c|}{5.3} & 5.0 \\ 
                          & 24                                                                     & \multicolumn{1}{c|}{9.7} & \multicolumn{1}{c|}{9.8} & \multicolumn{1}{c|}{5.3} & 5.0 \\ \hline
\end{tabular}
\caption{Emission reduction (in percentage) for hours-set for in-training fsFtS} \label{tab:hoursset4}
  \end{minipage}
\end{table}

It is evident that FtS tend to increase the reductions as the value of the hours-set increases in the flexible start version (see for example column in bold in Table \ref{tab:hoursset4}). This increment is justified by the fact that extending the time window by one day compared to the minimum duration of the workload allows for choosing the optimal starting time.

In general, the maximum average reduction is 16.3\% with a value of $24h$ for the hours-set (see values in bold in Table \ref{tab:hoursset3}). This value is achieved by the fsFtS with upstream data transfer, particularly by shorter workloads (SVM and AE). Averaged on every workload (so even including the longest one, i.e. HF-SCA), the reduction is 14.6\%. Changing the checking time does not lead to meaningful variations.

To have comprehensive and reliable comparison, Table~\ref{tab:redux} summarizes the results achieved by FS and PaR, averaged across all the available regions. Main takeaways from the table confirm the results obtained by Dodge et. al. \cite{dodge2022measuring}, they are the following:
\begin{itemize}
    \item The longest workload (HF-SCA) benefits the most from PaR, as the reduction in case of large time window is more than three times the case of small time window. At the same time, reduction grows very slowly in case of FS.
    \item The shorter workload (SVM) benefits the most from FS, as it doubles the emission reduction when the time window is large.
    \item In case of unconstrained training times, PaR is generally better than FS.
\end{itemize}

\begin{table}[]
\centering
\small
\begin{tabular}{|c|l|cccc|}
\hline
\multirow{2}{*}{Workload} & \multirow{2}{*}{Strategy} & \multicolumn{4}{c|}{Time window}                                                   \\
                          &                           & \multicolumn{1}{c|}{6}   & \multicolumn{1}{c|}{12}  & \multicolumn{1}{c|}{18}  & 24  \\ \hline
\multirow{2}{*}{IF}   & FS                        & \multicolumn{1}{c|}{3.9} & \multicolumn{1}{c|}{4.5} & \multicolumn{1}{c|}{5.2} & 6.1 \\
                          & PaR                       & \multicolumn{1}{c|}{3.2} & \multicolumn{1}{c|}{5.1} & \multicolumn{1}{c|}{5.7} & 7.0 \\ \hline

\multirow{2}{*}{SVM}       & FS                        & \multicolumn{1}{c|}{3.1} & \multicolumn{1}{c|}{5.2} & \multicolumn{1}{c|}{6.3} & 7.5 \\
                          & PaR                       & \multicolumn{1}{c|}{3.8} & \multicolumn{1}{c|}{5.4} & \multicolumn{1}{c|}{5.6} & 6.7 \\ \hline                          
\multirow{2}{*}{AE}      & FS                        & \multicolumn{1}{c|}{4.0} & \multicolumn{1}{c|}{4.9} & \multicolumn{1}{c|}{5.4} &6.6 \\
                          & PaR                       & \multicolumn{1}{c|}{3.2} & \multicolumn{1}{c|}{5.3} & \multicolumn{1}{c|}{6.0} & 7.3 \\ \hline
                                                
\multirow{2}{*}{HF-SCA}       & FS                        & \multicolumn{1}{c|}{2.3} & \multicolumn{1}{c|}{2.5} & \multicolumn{1}{c|}{2.6} & 2.8 \\
                          & PaR                       & \multicolumn{1}{c|}{1.5} & \multicolumn{1}{c|}{3.2} & \multicolumn{1}{c|}{4.9} & 5.0 \\ \hline
                          
\end{tabular}
    \caption{Emissions reduction (in percentage) across all regions for the two state-of-the-art strategies applied to the considered domain}
    \label{tab:redux}
\end{table}

\subsection{Evaluation} \label{sec:eval}
Considering the results just presented, several considerations can be made. First, it is evident that FtS tends to produce consistent results: as the values of the hours-set increase, the trends are generally increasing or at least stationary. This is an important aspect because, regardless of the workload duration, if there is no urgency to launch a training, choosing to provide a wider time window is certainly the best choice to reduce emissions. Next we report the main results evaluation for ssFtS and fsFtS.

\begin{itemize}
    \item \textit{Static-Start Follow-the-Sun.}
This version is ideal to use when there is the need to execute the training as soon as possible. Figure \ref{fig:timeplot} reports the start and execution times for the various strategies, for an arbitrary starting time, related to the IF workload. It can be observed that sometimes the considered strategies may postpone the starting time by $24h$, which is not the case with the ssFtS strategy.
More in general, looking at Table \ref{tab:delay} we can observe how much, on average, a training is delayed by the other strategies.
Despite striking a balance between emission reduction and time saving, ssFtS still achieves good results in terms of carbon savings. Particularly, our data reports carbon reductions in the range 5\% - 7\%, while state-of-the-art strategies remain within the range 2\% - 7\%, hence the results are slightly more solid.

\begin{figure}[t]
    \centering
    \includegraphics[width=1\linewidth]{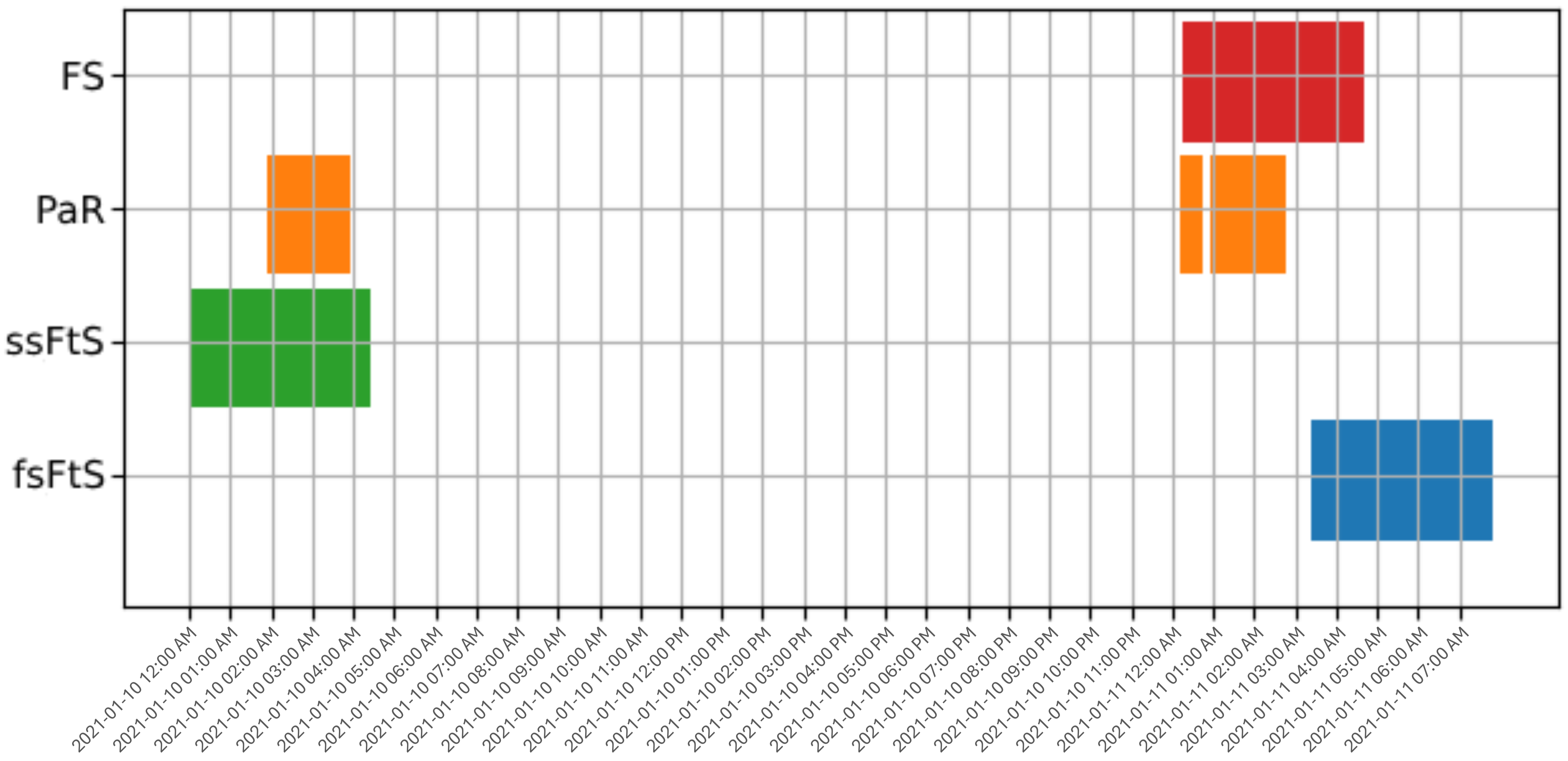}
    \caption{Timesheet for the Isolation Forest executed with different strategies. Coloured rectangles indicate the respective strategies running in the selected times.}
    \label{fig:timeplot}
\end{figure}
\begin{table}[t]
    \centering
    \small
    \begin{tabular}{|c|c|c|c|c|}
        \hline
        Strategy & IF & SVM & AE & HF-SCA\\
        \hline
        Pause and Resume & 17:34h & 15:50h & 16:57h & 21:33h\\
        \hline
        Flexible-Start & 13:25h & 12:45h & 12:56h & 19:15h\\
        \hline
        Flexible-Start FtS & 16:22h & 17:47h & 17:31h & 37:32h\\
        \hline
        No-Strategy execution & 4:15h & 2:30h & 3:30h & 16:00h\\
        \hline
    \end{tabular}
    \caption{Average durations for each workload applying the two state-of-the-art strategies and \textit{Flexible-Start FtS} with value from hours-set equals to $24h$, and $120m$ for checking-time. We do not report ssFtS as it has virtually no significant delay with respect to the no-strategy case.}
    \label{tab:delay}
\end{table}

\item \textit{Flexible-Start Follow-the-Sun.} As a complement to the static version, this strategy is preferable when there is flexibility in training timelines. Our results showcase a 16.3\% reduction in carbon emissions for shorter workloads and 10.0\% for the longer one. Specifically, similar to the static start case, the reduction increases as the hours-set increases. It should be noted that the values of the checking time have not a significant impact, or rather, they only have a visible contribution in terms of the average number of data transfers. This aspect brings an advantage: if there are no significant changes in reductions with an increase in the checking time, it means that the workload can be transferred a slightly lower number of times, minimizing the emissions associated with the transfer. The average number of data transfers ranges from 0 to 1.7, but this should not be misunderstood as the number of grid-region changes. For example, as seen in Figure \ref{fig:region_switches}, the HF-SCA workload is executed in two regions. Since this strategy is not constrained as in the static version, it has the ability to choose the best starting time. This translates into a benchmark that, on average, tends to be better than state-of-the-art strategies, which reach a maximum reduction of 7.5\% (Table \ref{tab:redux}).
Another crucial aspect and probably the most important is that it is not by chance that emission reduction tend to be higher for shorter workloads because -- as already reported -- the FS, under certain conditions, represents a lower bound for the fsFtS. This implies that this strategy can only improve the results of FS.

\begin{figure}[t]
    \centering
    \includegraphics[width=1.1\linewidth]{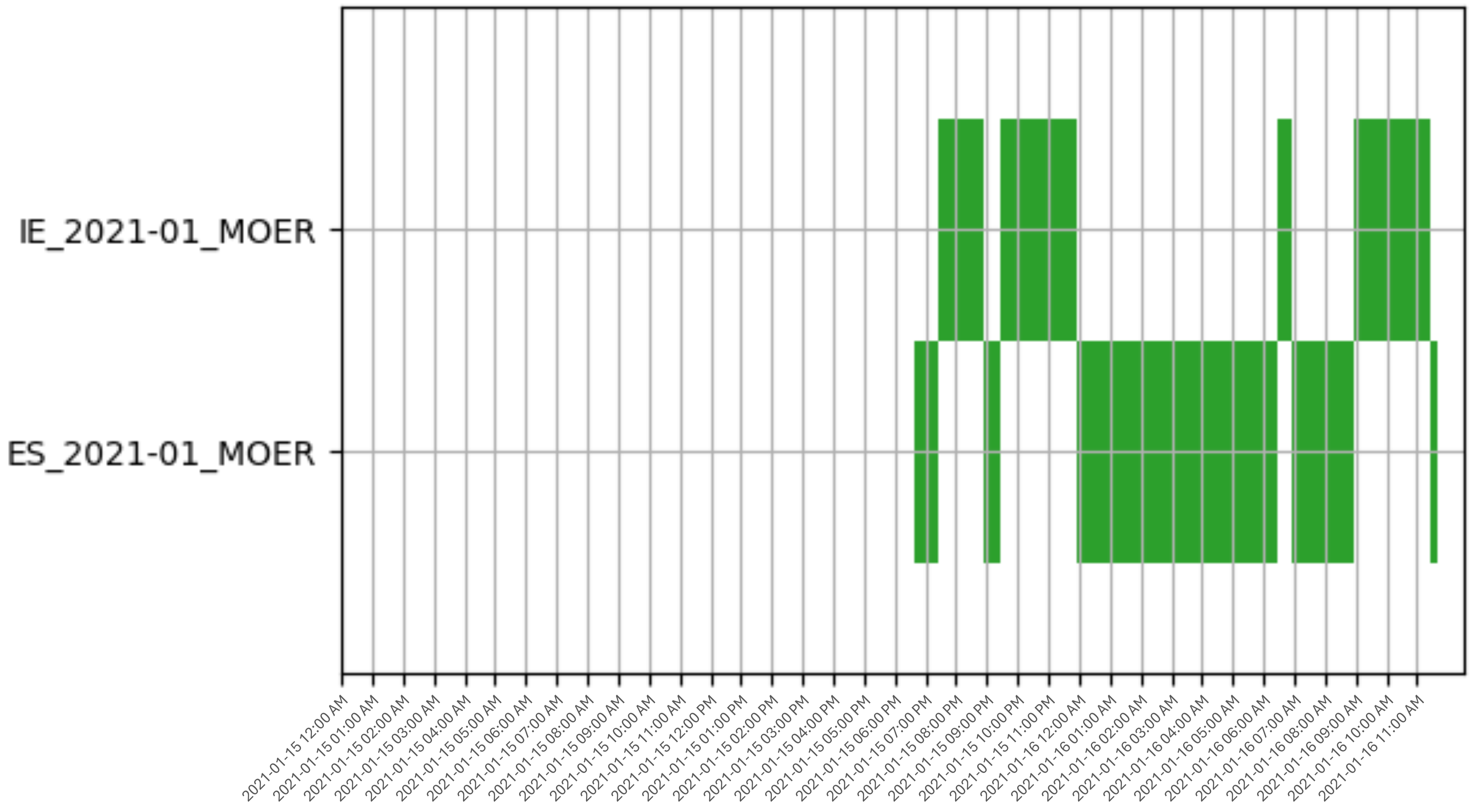}
    \caption{Region-switches for HF-SCA for \textit{Flexible-Start FtS}. IE\_2021-01\_MOER and ES\_2021-01\_MOER refer to Ireland and Spain.}
  \label{fig:region_switches}
\end{figure}
\end{itemize}

A final consideration can be made on the data transfer mode. The two transfer modes have shown to have very similar trends on average. We expected that the upstream approach would yield better results, and those expectations were met, although the difference is very small, if not negligible. Both the considered versions are optimized: in the case of upstream, it waits for the optimal moment before sending the data. Moreover, data already transferred to a grid-region is not sent again to the same region. This result is particularly interesting because it shows that the two modes, when optimized as in this case, tend to narrow the differences, allowing a user to choose either technique based on the available data, while still achieving very similar results.

\section{Discussion} \label{sec:discussion}
To fully understand the results, it is important to recall that PaR and FS represent the two state-of-the-art strategies implemented by Dodge et.al.~\cite{dodge2022measuring}. Comparative analyses of all strategies were conducted, and they were evaluated under the same conditions. In the evaluation section \ref{sec:eval}, we observed that the FtS strategy offers advantages on multiple fronts. 

First, we consider the reductions in carbon emissions by averaging the emissions for the entire year of 2021. However, the strategy object of investigation in this work not only operates horizontally in terms of time but also vertically in terms of regions. This distinction is important because the strength of FtS lies in its ability to optimize the very resource that state-of-the-art algorithms use to find a solution: time. By traversing different regions, this strategy -- at least in the ssFtS version -- has the potential to preserve 100\% of this resource, making it an additional element to consider in the discussion of the results.

Another crucial aspect arises from the fact that, to ensure valid results, we evaluated the performance of FS and PaR not only by averaging the results for an entire year but also across all regions. This detail highlights a third aspect, namely the dependence of the strategies on the reference region where the workload is executed.

\begin{table}[t]
    \centering
    \small
     \begin{tabular}{|c|c|c|c|}
        \hline
        
         Strategy & \begin{tabular}[c]{@{}c@{}}Avg time\\ dilatation\end{tabular} & \begin{tabular}[c]{@{}c@{}}Std time\\ dilatation\end{tabular} & \begin{tabular}[c]{@{}c@{}}Avg carbon\\ reduction\end{tabular} \\
        \hline
        Flexible-Start & 7:54h& 2:46h & 5.7\% \\
        \hline
        Pause and Resume & 11:15h& 3:26h & 6.5\% \\
        \hline
        
        Static-Start FtS & No dilatation & No dilatation & 5.9\% \\
        \hline

        Flexible-Start FtS & 15:36h& 3:24h & 14.6\% \\
        \hline
        
    \end{tabular}
    \caption{Quantitative comparison between strategies}
    \label{tab:over_time}
   
\end{table}

\paragraph{Time saving} In a context where time is a valuable resource for the industry, the intention to reduce carbon emissions with strategies that extend the training time can contradict the need to obtain a trained model as quickly as possible. Indeed, Table~\ref{tab:over_time} displays the average elongation in the duration across all workloads when subjected to state-of-the-art strategies. From the values, it is evident that the average time dilation for each workload resulting from the application of state-of-the-art strategies is around $11h$ for PaR, and nearly $8h$ for FS. On the other hand, ssFtS is not subject to such dilation, resulting in emission reductions that surpass those of FS. So this version is an alternative approach in case of constrained time, capable of reducing emissions through workload relocation. In addition to the benefits of reduced carbon emissions, it is important to consider the time saved, which translates into profitability for the industry. In fact, FS and PaR strategies are not designed to achieve results within a time window equal to the length of the training, unlike FtS.
\paragraph{Carbon emission reduction} If there is the possibility to choose a starting time within a time window, as is the case of state-of-the-art strategies, FtS can maximize its potential by combining the advantages of workload relocation with the benefits of selecting the best time instances. FtS optimizes time and serves as an alternative to FS. In fact, under certain conditions dictated by the size of the workload to be transferred, the  fsFtS has FS as its lower bound. Specifically, the dataset size must be such that the emissions resulting from its transfer do not exceed those that would be generated by staying in the current region. In this regard, the two optimized versions of data transfer help fulfil this condition. When this occurs, the results achieved by FtS in the worst-case scenario would be equivalent to those of FS. In Table~\ref{tab:over_time}, we can observe that the average reduction obtained by fsFtS (14.6\%) is the highest value. Moreover, we can confirm an aspect already observed by Dodge et.al.~\cite{dodge2022measuring}: FS is particularly efficient for short duration workloads. The same happens to fsFtS: average reduction is more than 16\% for shorter workloads, and 10\% for the longest one (see tables \ref{tab:hoursset3} and \ref{tab:hoursset4}). As Dodge et.al. already explained, shorter job is less subject to the variability of marginal emissions during the time window when we can choose the starting time.
    
\paragraph{Robustness with respect to the set of regions} To obtain reliable results we averaged the strategies' benefits over the entire year 2021 and across all regions, except for FtS strategies which always follows the same relocation path regardless of the launching region (Table \ref{tab:gdpr}). This makes FtS inherently robust with respect to the set of regions. In fact, the results of the PaR and FS strategies are highly dependent on the region they are located in, and if the region is not particularly green, it limits the potential for significant reductions.\\

\begin{table}[t]
    \centering
    \small
    \begin{tabular}{|c|c|c|c|}
        \hline
        Strategy & \begin{tabular}[c]{@{}c@{}}Robustness\\ w.r.t. regions\end{tabular} & \begin{tabular}[c]{@{}c@{}}Complex\\ architecture\end{tabular} & \begin{tabular}[c]{@{}c@{}}GDPR\\ limitations\end{tabular} \\
        \hline
        FS & No & No & No \\
        \hline
        PaR & No & No & No\\
        \hline
        ssFtS & Yes & Yes & Yes \\
        \hline
        fsFtS & Yes & Yes & Yes \\
        \hline
    \end{tabular}
    \caption{Qualitative comparison between strategies}
    \label{tab:gdpr}
\end{table}

The advantages brought by FtS occur on multiple fronts. FtS provides an additional choice to the decision maker because it not only guarantees promising reductions but also offers the possibility to prioritize slightly lower but still significant reductions while saving the entire available time. Such an option is not considered at all in state-of-the-art strategies. 

The obtained results can be achieved through the construction of an infrastructure capable of relocating a workload, which translates in relocation overhead and an increase in architectural complexity. This is a constraint that state-of-the-art strategies are not subject to (Table \ref{tab:gdpr}). Furthermore, precisely because state-of-the-art strategies do not involve relocation, they are exempt from the issues caused by regulations governing data exchange between countries, such as the European General Data Protection Regulation (GDPR).

Based on the achieved results, general considerations on the future adoption of FtS by the industry and the AI community can be outlined. What we have seen for FtS, that is also valid for the existing state-of-the-art strategies, is that the urgency in finding a more efficient balance between the usefulness of AI and its energy consumption compulsorily passes for a better match between workloads schedule and the availability of renewables. However, the avid rush towards the achievement of stunning performances is currently not regulated as in the case -- for instance --  of data protection. In fact, every company on the world is today free to burn all the carbon it wants in the name of accuracy. They should instead demonstrate concrete commitment in facing the climate challenge, or -- as goal 13 of UN's SDG recitate -- taking urgent action to combat climate change and its impacts. While we can discuss whether it is worth to implement FtS in that case or another, an harmonisation between AI and natural resources has to be enforced by regulators, because this work has proven that carbon-aware training strategies are effective in mitigating the carbon emissions of AI training, otherwise short-sighted business decisions will prevail.

Besides regulations' open issues, the integration of FtS in the development processes of companies involves a number of technical aspects that should be addressed. First, software developers today don't know whether the algorithms they use can be stopped or not, and if yes, what is the impact on the AI metrics of stopping and resuming jobs. To solve this issue, AI developers should facilitate the implementation of sustainable training strategies by clearly documenting their algorithm with the information on how to integrate their products within carbon-aware development processes. Second, current AI-Ops set of practices do not include the deploy of globally distributed workloads, nor existing tools like containers and orchestrators provide out-of-the-box solutions to facilitate the implementation of FtS. While many Cloud services have support for Kubernetes (available on AWS EKS, Azure AKS, and Google Kubernetes Engine), Kubernetes and similar container orchestration tools do not fully address the unique challenges of migrating AI workloads. Issues such as state management, data transfer, specialized hardware needs, latency, and compliance require additional planning and resources beyond what these tools offer.


\subsection{Threat to Validity}
The main threat to internal validity lies in the selection of the four AI algorithms, which may not be representative of the wide variety of algorithms available today. Additionally, focusing solely on the anomaly detection task, particularly on fraud detection, may limit the generalizability of our findings. To reduce this threat, we selected different types of architectures for the algorithms (both ML- and DL-based) with varying workload durations (ranging from $2:30h$ to $16:00h$).

The main threat to external validity stems from the partially synthetic scenario we defined in Section \ref{sec:cost}. Particularly, the assumption that the dataset, once compressed, does not have excessive dimensions allowed us to disregard transfer times and limit emissions related to networking. However, if this assumption does not hold in the real world, the emissions resulting from data transfer could make changing the grid-region not cost-effective. Moreover, longer transfer times imply a different carbon intensity at the target location compared to the initial check. To mitigate this threat, we suggest real-world implementations of FtS to check forecasted carbon intensity data at the destination within an interval that considers the transfer time.
Another threat to external validity is that the mere watching at lower carbon intensity may lead the more favourable computing centers to have a long queue of tasks waiting for execution, and the total carbon reduction may actually be lower.

The main threat to construct validity is the way we measure the energy consumption of data transfer. The literature suggests using the constant $0.023kWh/GB$~\cite{dodge2022measuring}. However, the metric should account for the distance between endpoints, link speed, the number of hops traversed, and their geographical locations. To mitigate this risk, we selected servers within the same geographical area (Europe), so the variance in their mutual distance is low. Despite the homogeneous distance, the carbon footprint of each grid-region varies a lot, due to different availability of renewable sources and differences in energy policies among the countries, making the experiment meaningful.

\section{Conclusions} \label{sec:conclusions}

In this work we have presented a comprehensive analysis of different strategies for reducing carbon emissions of AI training workloads in Cloud computing environments, particularly comparing the 'Follow-the-Sun' with those available in the state of the art.

First, we selected the field of Anomaly Detection (with a particular emphasis on fraud detection). This field was chosen to demonstrate that significant energy consumption is not a prerogative of heavier workloads like LLM or computer vision, but also because of the availability of a large dataset of banking transactions provided for an industrial use case. Four algorithms commonly used in this domain were identified, falling in the ML and DL fields, with varying lengths, sometimes similar, and in other cases significantly different. The idea is to have different workloads to conduct a more comprehensive analysis.

These trainings were executed on a GPU equipped machine, and the energy consumption was recorded every $5m$ to map them with CSV files containing marginal emission values for different regions, provided by WattTime, research partner in this work. A benchmark was conducted, comparing state-of-the-art strategies with the one proposed in this work, resulting in two significant findings: the FtS strategy not only achieves higher reductions (the best configuration hits an average of 14.6\% over all the workloads, 16.3\% for the shortest ones), but it also guarantees low emissions in situations where time is limited, preserving the minimum training duration.

For the sake of completeness, the algorithm that performed better among the considered four is HF-SCA, with an AUC score of 97\% (see Table \ref{tab:auc}). By the way, it is at the same time the one that emitted more CO$_{2}$: $1,600gCO_2eq$. However, its outstanding performance makes it the more suitable choice for the considered scenario. In general, in case of comparable performance, the greener algorithm should be preferred.

This work is intended to be useful for the following stakeholders:
\begin{itemize}
    \item Developers are stimulated to write more efficient code, while software architects  can decide whether to put effort in developing a complex globally distributed training architecture, or if it is better to focus on different phases of the AI pipeline like the inference phase. 
    \item Researchers in the AI field can formulate new strategies on the ground of the experience reported in this work and conduct new research to advance the field of green AI. Moreover, researchers in other fields of computer science could  apply the reported strategies to other heavy workloads.
    \item Organizations like companies or public bodies -- who are subjected to reporting carbon emissions and mitigation actions -- have a tool that aids the decision-making process and can demonstrate attention to the sustainability topic. They can choose the best strategy to improve the carbon footprint of their AI algorithms, according to workload length and re-training frequency.
    \item Financial stakeholders -- like banks (both private and national), fintech startups and regulatory bodies -- have a benchmark of the carbon emission of the most popular fraud detection AI algorithms and can suggest or implement actions to mitigate their impact.
\end{itemize}

We find some avenues for future research and improvements. First, our study focuses on a specific set of workloads and scenarios, and it would be valuable to extend the analysis to a wider range of applications and datasets, particularly with workloads that exceed $24h$ in duration. This would provide a more comprehensive understanding of the performance of the strategies in various contexts. Additionally, further investigation can be conducted to refine the decision- making process in selecting the optimal region for workload placement. Considering factors beyond carbon emissions, such as energy costs, network latency, and resource availability, could lead to more sophisticated and efficient strategies. In this work, a solution inspired by the Flexible Start was proposed with the intention of creating an even more efficient version of Follow-the-Sun. A future development could consider the idea of doing the same with the Pause and Resume strategy. Furthermore, exploring advanced optimization techniques, machine learning algorithms, or AI-based approaches could enhance the decision-making process and adapt it to dynamic and changing conditions. These methods could improve the accuracy and efficiency of workload placement decisions, leading to further reductions in carbon emissions.

\backmatter

\bmhead{Acknowledgements}
This work is the main result of the AMEDEA (Assessment and Mitigation of the Environmental impact of Deep lEarning Algorithms) project, which was supported by CINECA within the context of the ISCRA Class C programme (cod. IsCa7\_AMEDEA). Authors want to thank the research partner WattTime for providing the historical dataset of local-marginal emissions used for this work, and the Green Software Foundation both for the valuable research insights and for providing the conditions to establish the agreement with WattTime.

\section*{Declarations}
Not applicable.




\end{document}